\documentclass[aps,prl,twocolumn,superscriptaddress,showpacs, nobalancelastpage]{revtex4-1}
\usepackage{graphicx}
\usepackage{latexsym}
\usepackage{amsmath}
\usepackage{txfonts}
\usepackage{helvet}
\usepackage{braket}
\usepackage{amscd}

\usepackage{longtable}

\begin{document}

\title{Measurement of Untruncated Nuclear Spin Interactions via Zero- to Ultra-Low-Field Nuclear Magnetic Resonance} 

\author{J. W. Blanchard}
\affiliation{Materials Science Division, Lawrence Berkeley National Laboratory, Berkeley, CA, 94720}
\affiliation{Department of Chemistry, University of California at Berkeley, CA, 94720}
\affiliation{Helmholtz-Institut Mainz, Johannes Gutenberg University, Germany}
\author{T. F. Sjolander}
\affiliation{Materials Science Division, Lawrence Berkeley National Laboratory, Berkeley, CA, 94720}
\affiliation{Department of Chemistry, University of California at Berkeley, CA, 94720}
\author{J. P. King}
\affiliation{Materials Science Division, Lawrence Berkeley National Laboratory, Berkeley, CA, 94720}
\affiliation{Department of Chemistry, University of California at Berkeley, CA, 94720}
\author{M. P. Ledbetter}
\affiliation{Department of Physics, University of California at Berkeley, CA, 94720-7300}
\author{E. H. Levine}
\affiliation{Department of Chemistry, University of California at Berkeley, CA, 94720}
\author{V. S. Bajaj}
\affiliation{Materials Science Division, Lawrence Berkeley National Laboratory, Berkeley, CA, 94720}
\affiliation{Department of Chemistry, University of California at Berkeley, CA, 94720}
\author{D. Budker}
\affiliation{Helmholtz-Institut Mainz, Johannes Gutenberg University, Germany}
\affiliation{Department of Physics, University of California at Berkeley, CA, 94720-7300}
\affiliation{Nuclear Science Division, Lawrence Berkeley National Laboratory, Berkeley, CA, 94720}
\author{A. Pines}
\affiliation{Materials Science Division, Lawrence Berkeley National Laboratory, Berkeley, CA, 94720}
\affiliation{Department of Chemistry, University of California at Berkeley, CA, 94720}

\date{\today}

\begin{abstract} 

Zero- to ultra-low-field nuclear magnetic resonance (ZULF NMR) provides a new regime for the measurement of nuclear spin-spin interactions free from effects of large magnetic fields, such as truncation of terms that do not commute with the Zeeman Hamiltonian.
One such interaction, the magnetic dipole-dipole coupling, is a valuable source of spatial information in NMR, though many terms are unobservable in high-field NMR, and the coupling averages to zero under isotropic molecular tumbling.
Under partial alignment, this information is retained in the form of so-called residual dipolar couplings.
We report zero- to ultra-low-field NMR measurements of residual dipolar couplings in acetonitrile-2-$^{13}$C aligned in stretched polyvinyl acetate gels.
This represents the first investigation of dipolar couplings as a perturbation on the indirect spin-spin $J$-coupling in the absence of an applied magnetic field.
As a consequence of working at zero magnetic field, we observe terms of the dipole-dipole coupling Hamiltonian that are invisible in conventional high-field NMR.
This technique expands the capabilities of zero- to ultra-low-field NMR and has potential applications in precision measurement of subtle physical interactions, chemical analysis, and characterization of local mesoscale structure in materials.

\end{abstract}

\pacs{82.56.-b, 82.56.Dj, 76.60.-k, 76.60.Jx}

\keywords{}

\maketitle

Nuclear spin interactions are of substantial importance for many fields, including chemistry, quantum information processing, and precision measurement of fundamental symmetries. 
The most common technique for measuring such interactions is nuclear magnetic resonance (NMR), typically in large magnetic fields in order to maximize signal via higher nuclear spin polarization and sensitivity of inductive detection \cite{Slichter1990}. 
However, the only terms of the spin-coupling Hamiltonians that may be observed in high-field NMR are those that commute with the Zeeman Hamiltonian, which effectively truncates many interaction Hamiltonians that possess different symmetry. Recently, however, NMR experiments have been carried out in the opposite regime of very small magnetic fields \cite{Appelt2007, Appelt2010, Ledbetter2011, Colell2013}, taking advantage of advances in hyperpolarization \cite{Theis2011, Theis2012, Halse2008, Gloggler2011a} and new detection modalities \cite{Budker2007, Greenberg1998, Kominis2003, McDermott2002, Trabesinger2004, Allred2002, Schwindt2004}, which offer a significant time savings compared to earlier field-cycling techniques \cite{Ramsey1951, Zax1985}.
In zero- to ultra-low-field NMR (ZULF NMR), the strongest interactions are the local spin-spin couplings, which involve coupling tensors that are of different symmetry from the Zeeman Hamiltonian and are many orders of magnitude smaller in amplitude, thus permitting the direct observation of nuclear spin interactions that vanish at high magnetic fields.
Such terms that are only directly observable in the absence of large magnetic fields include the antisymmetric $J$-coupling (of importance for measurements of chirality and parity violation), several terms of the direct dipole-dipole coupling, and a number of as-yet-unobserved exotic interactions such as those mediated by pseudoscalar (axion-like) bosons \cite{Weinberg1978, Moody1984}, which would lead to anomalous spin-spin tensor couplings \cite{Ledbetter2013, Graham2013}, most of which do not commute with the Zeeman Hamiltonian.

As a proof of concept, we present in this Letter direct observation of the untruncated residual dipolar coupling between nuclear spins in a weakly aligning environment in the absence of an external magnetic field.
Dipolar couplings have long been used in high-field NMR to provide structural information in addition to the chemical shift.
Previous work demonstrated zero-field $J$-spectroscopy of several systems for chemical analysis \cite{Blanchard2013, Theis2013, Butler1}.
Additional information may also be obtained from zero-field NMR spectra via application of weak magnetic fields \cite{Ledbetter2011}.
In the regime where dipole-dipole interactions can be treated as a perturbation to the $J$-coupling, zero- to ultra-low-field (ZULF) NMR allows sensitive measurement of the dipole-dipole coupling tensor.
However, direct dipole-dipole couplings observed in solids are typically on the order of tens of kHz, substantially larger than $J$-couplings, and coherence and population lifetimes are often too short for current ZULF methodology.
Furthermore, all dipole-dipole coupling terms average to zero in isotropic liquids \cite{Slichter1990}.
Smaller, scaled couplings are obtained by weakly aligning the molecule of interest in anisotropic media, such as liquid crystals \cite{Diehl+Khetrapal, Emsley1975, Tjandra1997} or stretched gels \cite{Tycko2000, Chou2001}, where molecular motion is partially restricted, yielding residual couplings.
Such techniques have found widespread use in high-field NMR for structural measurements of proteins and small organic molecules \cite{Tjandra1997N, Clore1998, Prestegard2000, Trigo2012}.
Here, we investigate the effects of residual dipole-dipole couplings (RDCs) on the zero-field spectrum of a model XA$_3$ spin system: acetonitrile-2-$^{13}$C ($^{13}$CH$_3$CN, where we have found it valid to neglect the ${}^{14}$N spin due to self-decoupling arising from fast quadrupolar relaxation \cite{Spiess1977}) aligned in stretched crosslinked polyvinyl acetate (PVAc) gels.

\begin{figure}
  \includegraphics[width=3.0 in]{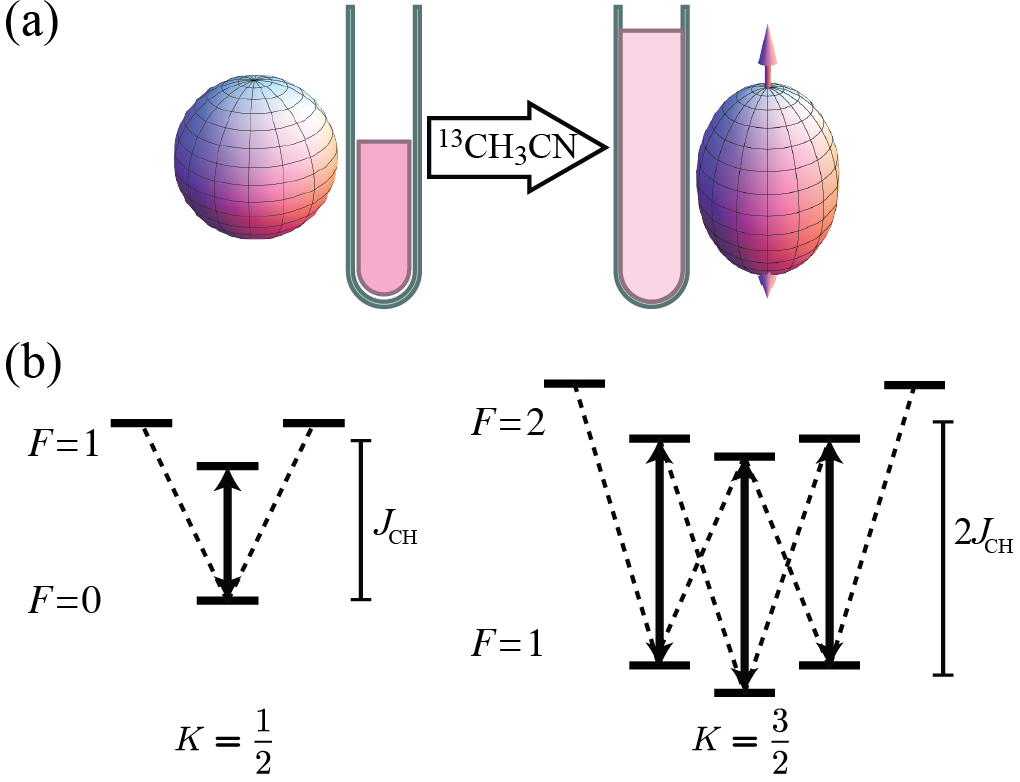}
  \caption{a) Schematic illustration of the change in symmetry that occurs during uniaxial stretching of the gel environment due to swelling with acetonitrile-2-$^{13}$C. The change in the order parameter is illustrated by the different three-dimensional shapes. b) Energy level structure of a partially ordered ${}^{13}$CH$_3$ spin system. $F$ is the total spin angular momentum, $K$ is the total proton angular momentum, and $J_{\rm CH}$ is the one-bond $^{13}$C -- $^1$H $J$-coupling. Solid arrows indicate allowed transitions, dashed lines indicate forbidden transitions. Note that for this system, both couplings are negative.}
  \label{Fig:cartoon}
\end{figure}

Polyvinyl acetate (PVAc) polymer sticks containing between 1-6\% v/v divinyl adipate (DVA) crosslinker were prepared in 5 mm NMR tubes as described in the Supporting Information. Anisotropic gels were prepared by adding acetonitrile to the tubes and allowing the polymers to swell for 2 weeks. Because the polymers were confined to the NMR tubes in which they were cast, swelling was uniaxial and equivalent to stretching along the axis of the NMR tube. A schematic representation of the process is shown in Fig.\ \ref{Fig:cartoon}(a).

In order to maximize the ZULF NMR signal, the samples were prepared using labeled acetonitrile-2-${}^{13}$C to which was added 5\% v/v deuterated acetonitrile for the purpose of high-field NMR characterization. The molecular order parameter \cite{Saupe1964, Saupe1965} for acetonitrile in the stretched gel environment was determined by analyzing the quadrupolar splitting of the deuterium resonance \cite{Diehl+Khetrapal, Caspary1969} using a 14.1 T NMR spectrometer with deuterium frequency 92.1 MHz (for additional details, see Supporting Information). The value for the electric-field gradient around the deuterium nuclei in acetonitrile was obtained from the literature \cite{Caspary1969}. 

\begin{figure*}
  \includegraphics[width=7.0 in]{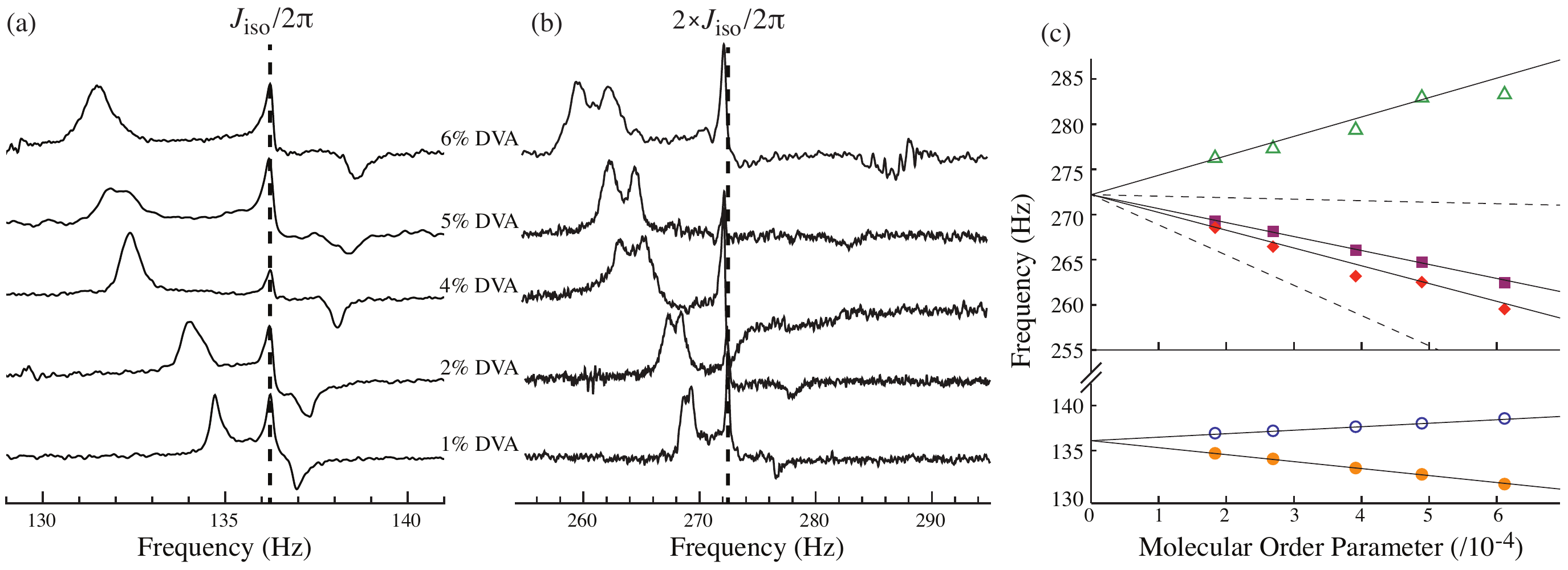}
  \caption{Zero-field spectra of acetonitrile-2-$^{13}$C with different degrees of ordering arising from the concentration of the cross-linker divinyl adipate (DVA). (a) $K=\frac{1}{2}$ and (b) $K=\frac{3}{2}$ peaks. (c) Peak positions as a function of molecular order parameter. The lines are calculated transition frequencies from Eqs. (\ref{eq:1Jallowed}-\ref{eq:2Jforbidden3}) with no free parameters. The order parameter for each sample was calculated from high-field deuterium quadrupole splittings using literature values for the electric field gradient \cite{Caspary1969} and the dipole-dipole coupling strengths were calculated from Eq. (\ref{eq:Dorderparameter}) using bond lengths from Ref. \cite{Kessler1950} . Solid symbols represent allowed $\Delta m_F=0$ transitions, open symbols represent transitions with $\Delta m_F = \pm 1$. Dashed lines indicate possible transitions that are not resolved.}
  \label{Fig:lightshiftspectra}
\end{figure*}

\begin{figure}
  \includegraphics[width=3.4 in]{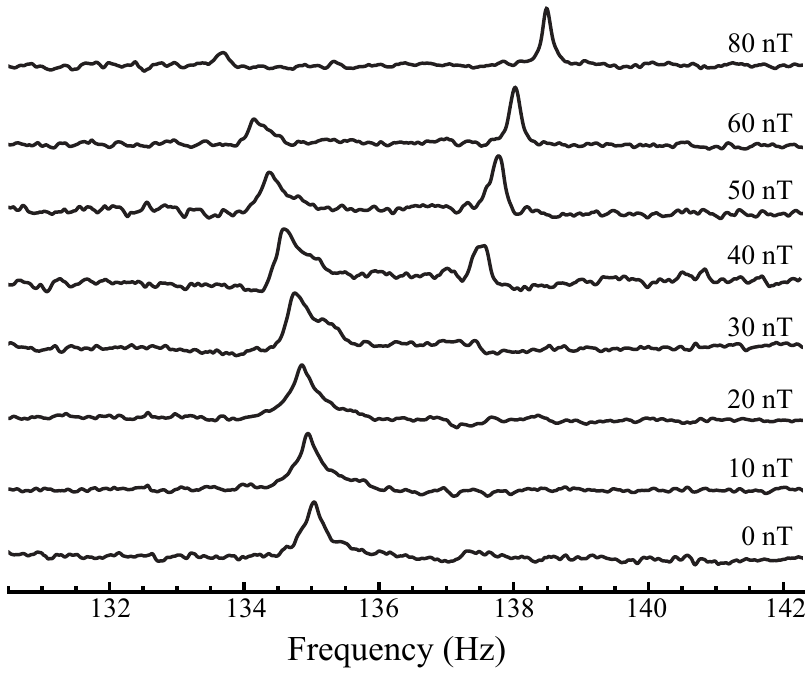}
  \caption{Acetonitrile-2-$^{13}$C $K=\frac{1}{2}$ peaks as a function of magnetic field applied orthogonal to the director axis, collinear with the magnetometer pump beam.
  }
  \label{Fig:fields}
\end{figure}

The ZULF NMR apparatus has been described previously \cite{Ledbetter2009, Theis2011, Ledbetter2011}. Samples were pre-polarized in a 2~T permanent magnet located outside of the magnetic shielding for $\sim$20~s and then shuttled pneumatically to the zero-field region over 0.5 -- 1~s. NMR signals were detected with an atomic magnetometer featuring a 0.6 $\times$ 0.6 $\times$ 1.0 cm$^3$ $^{87}$Rb vapor cell operating at 180$^\circ$C. Transient signals were collected over $\sim$20~s. The spectra in Fig.\ \ref{Fig:lightshiftspectra} are the average of between 256 and 1024 transients, and the spectra in Fig.\ \ref{Fig:fields} are the average of 8 transients.

The spin Hamiltonian in the presence of $J$-couplings and dipole-dipole interactions is
\begin{multline}
\mathcal{H} = \hbar \sum_{j;k>j} J_{jk} \mathbf{I}_j \cdot \mathbf{I}_k \\
- \hbar^2 \frac{\mu_0}{4\pi} \sum_{j;k>j} \frac{\gamma_j \gamma_k }{r_{jk}^3} \left[ 3 \left( \mathbf{I}_j \cdot \hat{\mathbf{r}}_{jk} \right) \left( \mathbf{I}_k \cdot \hat{\mathbf{r}}_{jk} \right) - \mathbf{I}_j \cdot \mathbf{I}_k \right],
\label{Eq.Htotal}
\end{multline}
where $\hbar$ is the reduced Planck constant, $\mu_0$ is the vacuum permeability, $\gamma_j$ and $\gamma_k$ are the gyromagnetic ratios of spins $\mathbf{I}_j$ and $\mathbf{I}_k$, and $\mathbf{r}_{jk}$ is the internuclear vector connecting the spins.
In the case of isotropic liquids, the dipole-dipole interaction term averages to zero.
However, in aligned samples, such as stretched gels, the motional averaging of the dipole-dipole term is incomplete.
For the system studied here, the swelling of the polymer gel with acetonitrile along the axis of the NMR tube leads to an orientational probability distribution of the solvent molecules that is slightly anisotropic, with the preferential alignment axis (the director) determined by the swelling direction \cite{Tycko2000, Chou2001}.
This axis is collinear with the sensitive direction of the detector, and is denoted $z$.
Because of the rapid rotation of the acetonitrile methyl group and the axial symmetry of the alignment medium, the $x$ and $y$ components of the $\mathbf{r}_{jk}$ vectors are averaged to zero, and the $z$ components are scaled by the degree of alignment.
Considering these averaging effects on the second term of Eq. (\ref{Eq.Htotal}), the residual dipolar coupling Hamiltonian is
\begin{equation}
\mathcal{H}_{RDC} = - \hbar \sum_{j;k>j} {D}_{jk}  \left(3 I_{j,z} I_{k,z} - \mathbf{I}_j \cdot \mathbf{I}_k \right),
\label{Eq.RDC}
\end{equation}
where
\begin{equation}
{D}_{jk} = \frac{\mu_0}{4\pi}\frac{\gamma_j \gamma_k \hbar}{r_{jk}^3} \frac{1}{2}\left\langle3 \cos^2 \theta_{jk}-1 \right\rangle,
\label{eq:D}
\end{equation}
and where $\theta_{jk}$ is the angle between the internuclear vector and the $C_3$-axis of acetonitrile.
We may also define the coupling as being directly proportional to the molecular order parameter $\mathcal{S}_{zz}=\frac{1}{2}\left\langle3 \cos^2 \theta_{z}-1 \right\rangle$ \cite{Saupe1964, Saupe1965}, where $\theta_{z}$ is the angle between the $C_3$-axis of acetonitrile and the laboratory $z$-axis and the time average provides a measure of the extent of alignment between the two axes systems, such that
\begin{equation}
{D}_{jk} = \frac{\mu_0}{4\pi}\frac{\gamma_j \gamma_k \hbar}{r_{jk}^3} \frac{1}{2}\left(3 \cos^2 \phi_{jk}-1 \right) \mathcal{S}_{zz},
\label{eq:Dorderparameter}
\end{equation}
where $\phi_{jk}$ is the angle between $\mathbf{r}_{jk}$ and the $C_3$ axis, $\pi/2$ for $D_{\rm HH}$ and the tetrahedral angle ($2 \arctan\sqrt{2}$) for $D_{\rm CH}$.
Because the order parameter $\mathcal{S}_{zz}$ may be readily extracted from high-field deuterium NMR spectra, the only remaining parameters for the dipole-dipole coupling strengths are the angles and distances that define the molecular geometry, which may be obtained from the literature \cite{Kessler1950}.

It is worth pointing out the difference between the effective residual dipole coupling Hamiltonian in zero field and in high field.
The zero-field heteronuclear coupling term in terms of the total $^{1}$H angular momentum $\mathbf{K}$ and the $^{13}$C angular momentum $\mathbf{S}$ is
\begin{equation}
\mathcal{H}^{(het)} = - \hbar {D}_{\rm CH} \left(3 K_z S_z - \mathbf{K} \cdot \mathbf{S} \right),
\label{ZFdipole}
\end{equation}
as compared to the high-field case,
\begin{equation}
\mathcal{H}^{(trunc)} = - 2\hbar {D}_{\rm CH} K_z S_z,
\label{Eq.Htrunc} 
\end{equation}
wherein a term of the form
\begin{equation}
\mathcal{H}_{RDC} -\mathcal{H}^{(trunc)} =\frac{\hbar}{2} D_{\mathrm{CH}} \left(K_+S_-+K_-S_+\right),
\label{Eq.Hflipflop}
\end{equation}
is truncated because it does not commute with the high-field Zeeman Hamiltonian. The zero-field Hamiltonian is untruncated, and thus includes this so-called ``heteronuclear flip-flop'' term that is invisible to high-field NMR.

In the regime where ${D}_{jk} \ll {J}_{jk}$, the residual dipolar couplings may be treated as a perturbation on the $J$-coupling, yielding energy shifts that are calculated in the Supporting Information, and summarized schematically in Fig.\ \ref{Fig:cartoon}(b), along with the allowed transitions.

Because the observable in this experiment is the $z$-magnetization $M_z(t)\propto \rm{Tr}\ \{\it \rho(t) \sum_j I_{j z}\gamma_j\}$, the  detectable coherences are those with $\Delta F=0, \pm1$ and $\Delta m_F=0$.
An additional selection rule, $\Delta K=0$, arises in the case of equivalent spins (e.g. the methyl protons in acetonitrile) because $\mathbf{K}^2$ commutes with the Hamiltonian \cite{Ledbetter2009, Ledbetter2011}.
It follows that there is one allowed transition between $K=\frac{1}{2}$ states, between the $|F=0,m_F=0\rangle$ and $|F=1,m_F=0\rangle$ states. Based on the first-order energy shifts presented in the Supporting Information, this transition has frequency
\begin{equation}
\omega_{0,0}^{1,0}= J_{\mathrm{CH}} + D_{CH}.
\label{eq:1Jallowed}
\end{equation}
In addition, there are nominally forbidden (assuming that the detector is only sensitive in the $z$-direction) $\Delta m_F=\pm1$ transitions with frequency
\begin{equation}
\omega_{0,0}^{1,\pm1}= J_{\mathrm{CH}} - \frac{D_{CH}}{2}.
\label{eq:1Jforbidden}
\end{equation}

For the transitions between $K=\frac{3}{2}$ states, there are allowed transitions with frequencies
\begin{align}
\label{eq:2Jallowed1} \omega_{1,0}^{2,0}&= 2J_{\mathrm{CH}} + 2D_{CH},\\
\omega_{1,\pm1}^{2,\pm1}&= 2J_{\mathrm{CH}} +\frac{1}{2}\left(D_{CH}+3D_{HH}\right),
\label{eq:2Jallowed2}
\end{align}
and nominally forbidden transitions $\Delta m_F=\pm1$ with frequencies
\begin{align}
\label{eq:2Jforbidden1} \omega_{1,\pm1}^{2,\pm2}&= 2J_{\mathrm{CH}} -\frac{1}{4}\left(7D_{CH}+3D_{HH}\right),\\
\label{eq:2Jforbidden2} \omega_{1,0}^{2,\pm1}&= 2J_{\mathrm{CH}} +\frac{1}{4}\left(5D_{CH}-3D_{HH}\right),\\
\omega_{1,\pm1}^{2,0}&= 2J_{\mathrm{CH}} +\frac{1}{4}\left(5D_{CH}+9D_{HH}\right).
\label{eq:2Jforbidden3}
\end{align}

If the detector axis is not exactly aligned with the director/quantization axis, the nominally forbidden transitions become observable.

Zero-field spectra of acetonitrile-2-$^{13}$C ($^{13}$CH$_3$CN) in stretched polyvinyl acetate gels are shown in Fig. \ref{Fig:lightshiftspectra}(a) for increasing values of the molecular order parameter.
As the order parameter increases, the $K=\frac{1}{2}$ peaks corresponding to the ordered portion of the sample split, while the $K=\frac{1}{2}$ peak corresponding to excess isotropic liquid external to the gel remains unchanged.
The lower-frequency peak in Fig. \ref{Fig:lightshiftspectra}(a) corresponds to the $\Delta m_F=0$ transition described by Eq.\ (\ref{eq:1Jallowed}) and the higher-frequency peak corresponds to the $\Delta m_F=\pm1$ transition described by Eq.\ (\ref{eq:1Jforbidden}).
The magnitude and phase of the $\Delta m_F=\pm1$ peak are determined by the projection of the initial spin-state population onto the transverse component of the detection operator, and is thus a signature of imperfections in the experimental configuration.
Because the $\Delta m_F=\pm1$ peaks are consistently narrower than the $\Delta m_F=0$ peaks in Fig.~\ref{Fig:lightshiftspectra}(a), it appears that the linewidth is affected by inhomogeneity in the gel producing a distribution of order parameters and thus a spread in transition frequencies proportional to $D_{\rm CH}$.

Figure \ref{Fig:lightshiftspectra}(b) shows four $K=\frac{3}{2}$ peaks, three from the aligned acetonitrile-2-$^{13}$C, and one from the isotropic liquid. The two lower-frequency peaks arise from the $\Delta m_F=0$ transitions described by Eqs.\ (\ref{eq:2Jallowed1}-\ref{eq:2Jallowed2}) and the small higher-frequency peak corresponds to the $\Delta m_F=\pm1$ transition described by Eq.\ (\ref{eq:2Jforbidden1}). Transitions corresponding to Eqs.\ (\ref{eq:2Jforbidden2}-\ref{eq:2Jforbidden3}) are not resolved.

In high-field NMR, terms in the Hamiltonian that do not commute with the Zeeman Hamiltonian are neglected, due to their immeasurably small effect on the NMR spectrum.
Due to the absence of a large Zeeman interaction, ZULF NMR provides spectroscopic access to all spin-coupling terms \cite{Zax1984}.
In traditional high-field NMR, only part of the heteronuclear dipolar coupling, $\mathcal{H}^{(trunc)}$, from Eq.~(\ref{Eq.Htrunc}), yields measurable effects in the spectrum.
By itself, this term would yield no shift of the $|F=0,m_F=0\rangle \leftrightarrow |F=1,m_F=0\rangle$ transition.
In Fig.\ \ref{Fig:lightshiftspectra}(a), the residual dipolar couplings clearly shift the peak relative to the isotropic liquid showing the observation of the untruncated Hamiltonian of Eq.~(\ref{Eq.RDC}), including the contribution of Eq.~(\ref{Eq.Hflipflop}), an interaction ``invisible'' to traditional NMR.
The absence of truncation that permits observation of this term has also been demonstrated via the preparation of heteronuclear spin-singlet states in Ref.\ \cite{Emondts2014}.

As shown in Fig.~\ref{Fig:lightshiftspectra}(c), the frequency shift varies linearly with the order parameter. The data closely match the simulated curves, which are calculated from the order parameter using Eqs.~(\ref{eq:Dorderparameter}) and (\ref{eq:1Jallowed}-\ref{eq:2Jforbidden3}).

In the situation where the sensitive axis of the magnetometer is parallel to the director of the gel alignment, only the $\Delta m_F = 0$ transitions are detected.
If the detection axis (or direction of alignment) is rotated, the $\Delta m_F = \pm 1$ transitions are also observable, leading to peaks at higher frequency, with intensity dependent on the initial spin state populations and the angle between the detection and alignment axes.
The additional peaks appear in Fig.~\ref{Fig:lightshiftspectra} because the measurements were carried out using a magnetometer configuration that featured a rotated axis of detection due to a non-zero effective field at the Rb cell. 
We attribute this effect to imperfections in the magnetometer configuration, potentially related to laser alignment, AC Stark shifts, or a combination of the two.
We point out that the effect was diminished after expanding the pump beam (thus decreasing the laser power density) and subsequently realigning the optics for the experiments in Fig.~\ref{Fig:fields}.
It is worth noting that the rotation of the detection axis is necessary for the detection of the higher-frequency peaks only due to the axial symmetry of the system under study, which causes the terms of the dipolar coupling Hamiltonian that depend on the azimuthal angle to be averaged to zero.
If these terms are not averaged to zero (e.g. in a biaxial phase), they will further lift degeneracy and yield additional peaks \cite{Thayer1987LC}.

We have also investigated the effect of applied magnetic fields on the spectrum, as shown in Fig.\ \ref{Fig:fields}.
When the effective detection operator is collinear with the gel director axis, only the $\Delta m_F = 0$ transition is observed (corresponding here to a 10 nT applied field).
As the field is increased, however, rotation of the nuclear spins mixes states with different $m_F$, and rotation of the alkali electron spins elicits a change in the sensitive axis of the detector.
The overall result is that as the field is increased, the vectors defining the detection operator and the quantization axis cease to be collinear.
This in turn leads to the observation of $\Delta m_F = \pm1$ transitions, which become dominant above 50 nT, at which point the effective detection operator has been rotated substantially away from the director axis.

In the regime where the Zeeman interaction strength is on the order of the residual dipolar coupling, the peak frequencies in Fig.~\ref{Fig:fields} do not vary linearly with the applied field strength. This is because the dipole-dipole coupling Hamiltonian does not commute with the Zeeman Hamiltonian, and thus first-order perturbation is no longer sufficient to describe the system.

In conclusion, we have demonstrated the direct influence of the heteronuclear dipole-dipole coupling ``flip-flop'' term (which is not directly observable in the high-field regime) on ZULF NMR spectra.
The results are in agreement with a zero-free-parameter model of residual dipolar coupling utilizing literature values for internuclear distances and the electric field gradient of acetonitrile, the latter being used to determine the order parameter via high-field deuterium NMR.
The sub-Hz resolution of ZULF NMR may be of use for chemical and structural analysis of small molecules, as well as for precision measurement searches for anomalous spin-spin couplings along the lines of Ref. \cite{Ledbetter2013}.

In principle, all terms of the spin coupling Hamiltonian are observable in ZULF NMR, increasing the information available in NMR spectra.
With appropriate systems (e.g. aligned chiral molecules) and continuing improvements in polarization and magnetometer sensitivity, ZULF NMR may be a promising method to measure subtle interactions such as the as-yet-unobserved antisymmetric components of the $J$-coupling tensor \cite{Harris2009}.
Based on calculations in Ref.~\cite{Barra1996}, measurement of a non-zero antisymmetric $J$-coupling could permit the observation of a first-order energy shift arising from parity non-conservation in the molecular Hamiltonian.

This technique may also find application as a probe of material structure, allowing for the measurement of interactions with lower than azimuthal symmetry. For example, measurement of the full anisotropic spin-spin coupling tensor of small molecular probes within porous materials may provide a more complete description of local geometry than do measurements of total surface area or average pore diameter.

\section{Acknowledgement}

This work was supported in part by the Director, Office of Science, Office of Basic Energy Sciences, Materials Sciences and Engineering Division, of the US Department of Energy under Contract No. DE-AC02-05CH11231 (infrastructure for atomic magnetometry detection) and by the National Science Foundation under award \#CHE-1308381 (zero-field NMR analytical chemistry and spectroscopy). J.W.B. was also supported by a National Science Foundation Graduate Research Fellowship under Grant No. DGE-1106400. We would like to thank Roberto R. Gil for helpful discussions of stretched gel preparation.

\bibliography{ZULF_RDCs}

\end{document}